\documentstyle{mn2e}
\input epsf
\def\plotone#1{\centering \leavevmode
\epsfxsize=\columnwidth \epsfbox{#1}}
\def\plotwide#1{\centering \leavevmode
\epsfxsize=1.98\columnwidth \epsfbox{#1}}
\def\plottwo#1#2{\centering \leavevmode
\epsfxsize=.99\columnwidth \epsfbox{#1} \hfil
\epsfxsize=.99\columnwidth \epsfbox{#2}}

\newcommand{\beq}{\begin{equation}}
\newcommand{\eeq}{\end{equation}}
\newcommand{\beqa}{\begin{eqnarray}}
\newcommand{\eeqa}{\end{eqnarray}}

\def\deg{$^{\circ}$}

\title[Positron annihilation spectrum from the Galactic Center
region]{Positron annihilation spectrum from the Galactic Center region
observed by SPI/INTEGRAL}

\author[Churazov et al.]{E.~Churazov,$^{1,2}$ R.~Sunyaev,$^{1,2}$
S.~Sazonov,$^{1,2}$ M.~Revnivtsev,$^{1,2}$ D.~Varshalovich,$^{3}$ \\
$^1$ Max-Planck-Institut f\"ur Astrophysik, Karl-Schwarzschild-Strasse 1, 85741
Garching, Germany\\
$^2$ Space Research Institute (IKI), Profsoyuznaya 84/32, Moscow 117997, 
Russia\\
$^3$ Ioffe Institute, Polytekhnicheskaya 26, St Petersburg 194021, Russia
}

\pagerange{\pageref{firstpage}--\pageref{lastpage}}
\pubyear{2003}

\begin{document}
\maketitle

\label{firstpage}
\begin{abstract}
The electron-positron annihilation spectrum observed by SPI/INTEGRAL
during deep Galactic Center region exposure is reported. The line
energy (510.954$\pm$0.075 keV) is consistent with the unshifted
annihilation line.  The width of the annihilation line is
2.37$\pm$0.25 keV (FWHM), while the strength of the ortho-positronium
continuum suggests that the dominant fraction of positrons (94$\pm$6\%) form
positronium before annihilation.  Compared to the previous missions
these deep INTEGRAL observations provide the most stringent
constraints on the line energy and width.

Under the assumption of an annihilation in a single-phase medium these
spectral parameters can be explained by a warm $T_e\sim 7000-4~10^4$
K gas with the degree of ionization larger than a few $10^{-2}$. One of the
wide-spread ISM phases - warm ($T_e \sim 8000$ K) and weakly ionized
(degree of ionization $\sim$ 0.1) medium satisfies these criteria.
Other single-phase solutions are also formally allowed by the data
(e.g. cold, but substantially ionized ISM), but such solutions are
believed to be astrophysically unimportant.

The observed spectrum can also be explained by the annihilation in a
multi-phase ISM. The fraction of positrons annihilating in a very hot
($T_e \ge 10^6$ K) phase is constrained to be less than $\sim$8\%.
Neither a moderately hot ($T_e \ge 10^5$ K) ionized medium nor a very cold
($T_e \le 10^3$ K) neutral medium can make a dominant contribution to
the observed annihilation spectrum. However, a combination of
cold/neutral, warm/neutral and warm/ionized phases in comparable 
proportions could also be consistent with the data.
\end{abstract}


%

\sloppypar

\section{Introduction}
The annihilation line of positrons at 511 keV is the brightest gamma-ray
line in the Galaxy. First observed with a NaI scintillator as a $\sim$ 476
keV line coming from the Galactic Center (GC) region (Johnson, Harnden
\& Haymes, 1972; Johnston \& Haymes, 1973), it was subsequently
unambiguously identified with a narrow ($FWHM<3.2$ keV)~~ $e^+e^-$
annihilation line using germanium detectors (Leventhal, MacCallum,
Stang, 1978). Since then many balloon flights and several space
missions have measured the spatial distribution and spectral
properties of the line. A summary of the high energy resolution
observations of the 511 keV line prior to INTEGRAL and the first
SPI/INTEGRAL results can be found in Jean et al. (2003) and Teegarden
et al. (2004).

\begin{figure} 
\plotone{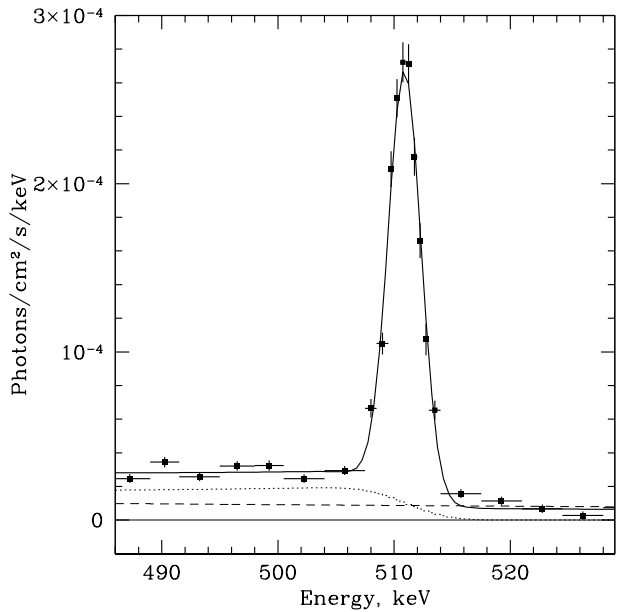}
\caption{Spectrum of the $e^+e^-$ annihilation radiation (fixed
background model) detected by SPI from the GC region and the best fit
model (thick solid line, see Table \ref{tab:fit} for parameters).  The
dotted line shows the ortho-positronium radiation and the dashed line
shows the underlying power law continuum.
\label{fig:spraw}
}
\end{figure}

Positrons in the Galaxy can be generated by a number of processes,
including e.g. radioactive $\beta^+$ decay of unstable isotopes
produced by stars and supernovae, jets and outflows from
the compact objects, cosmic rays interaction with the interstellar
medium (ISM), and annihilation or decay of dark matter particles. An
important problem is to determine the total $e^+e^-$ annihilation rate
in the Galaxy and to accurately measure the spatial distribution of
the annihilation radiation. This is a key step in determining the
nature of the positron sources in the Galaxy. Another problem is to
measure the annihilation spectrum including the 511 keV
line itself and the $3\gamma$ continuum arising from the decay of
ortho-positronium. This information reveals the properties of the ISM
where positrons are annihilating.

Here we concentrate on the latter problem and report below the
measurements of the $e^+e^-$ annihilation spectrum (including
$3\gamma$ continuum) based on SPI/INTEGRAL observations of the GC
region over the period from Feb., 2003 through Nov., 2003. The core of
the data set is a deep 2 Msec GC observation, carried out as part of
the Russian Academy of Sciences share in the INTEGRAL data. Previously
reported results on the 511 keV line shape (Jean et al. 2003) are
based on a significantly shorter data set. We use here a completely
independent package of SPI data analysis and for the first time report
the results on the ortho-positronium continuum measurements based on
the SPI data (Fig.\ref{fig:spraw}). The imaging results will be
reported elsewhere.

The structure of the paper is as follows. In Section 2 we describe the
data set and basic calibration procedures. Section 3 deals with the
spectra extraction. In Section 4 we present the basic results of
spectral fitting. In Section 5 we discuss constraints on the
annihilation medium. The last section summarizes our findings.

\section{Observations and data analysis}
SPI is a coded mask germanium spectrometer on board INTEGRAL (Winkler
et al., 2003), launched in October 2002 aboard a PROTON rocket. The
instrument consists of 19 individual Ge detectors, has a field of view
of $\sim$16\deg (fully-coded), an effective area of $\sim 70~cm^2$ and
the energy resolution of $\sim$2 keV at 511 keV (Vedrenne et al.,
2003, Attie et al., 2003). Good energy resolution makes SPI an
appropriate instrument for studying the $e^+e^-$ annihilation line.

\subsection{Data set and data selection}
A typical INTEGRAL observation consists of a series of pointings, during
which the main axis of the telescope steps through a 5x5 grid on the
sky around the 
position of the source. Each individual pointing usually lasts s few
ksec. A detailed description of the dithering patterns is given by
Winkler et al. (2003). For our analysis we use all data available to
us, including public data, some proprietary data (in particular,
proposals 0120213, 0120134) and the data available to us through the
INTEGRAL Science Working Team. All data were taken by SPI during the
period from Feb., 2003 through Nov., 2003. The choice of this time
window was motivated by the desire to have as uniform a data set as
possible. The first data used are taken immediately after the first SPI
annealing, while the last data used were taken prior to the failure of
one of the 19 detectors of SPI. While analysis of the GC data taken
after Nov. 2003 is possible, the amount of data (in public access)
which can be used for background modeling is at present limited.

Prior to actual data analysis all individual observations were
screened for periods of very high particle background. We use the SPI
anticoincidence (ACS) shield rate as a main indicator of high
background and dropped all observations with an ACS rate in excess of 3800
cnts/s. Several additional observations were also omitted from the
analysis, e.g. those taken during cooling of SPI after the annealing
procedure.

For our analysis we used only single and PSD events and when available
we used consolidated data provided by the INTEGRAL Science Data Center
(ISDC, Courvoisier et al, 2003).

\subsection{Energy calibration}
\begin{figure} 
\plotone{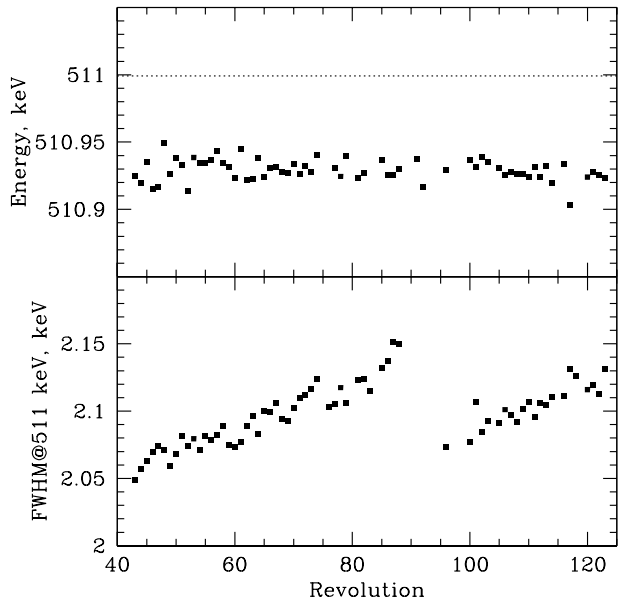}
\caption{The upper panel shows the energy of the background 511 keV line
vs the revolution number. With the assumed linear channel/energy relation
the line is shifted from the true energy of the $e^+e^-$ annihilation line
by $\sim$ 0.07 keV, the RMS deviation from the mean energy is only
0.0078 keV.  The bottom panel shows the intrinsic energy resolution of
SPI at the 511 keV line evaluated from fitting two narrow background lines
at 438 and 584 keV. Statistical errors are omitted for clarity.
\label{fig:ecal}
}
\end{figure}

As a first step all observations have been reduced to the same
gain. Trying to keep the procedure as robust as possible we assume
a linear relation between detector channels and energies and use four
prominent background lines (Ge$^{71}$ at 198.4 keV; Zn$^{69}$ at
438.6; Ge$^{69}$ at 584.5 keV and Ge$^{69}$ at 882.5 keV, see
Weidenspointner et al., 2003 for the comprehensive list of SPI
background lines) to determine the gain and shift for each
revolution. While the linear relation may not be sufficient to provide
the absolute energy calibration to an accuracy much higher than 0.1 keV
over the SPI broad energy band, the relative accuracy is high (see
Fig.\ref{fig:ecal}). Shown in the top panel is the energy of the
background 511 keV line as a function of the revolution number. While
the deviation from the true energy of the $e^+e^-$ line is $\sim$ 0.07
keV, the RMS deviation from the mean energy is only 0.0078 keV.  The
best fit energy of the background line for the combined spectrum of all
SPI observations within 30\deg of GC is 510.938 keV, compared to the
electron rest energy of 510.999 keV. The energies quoted below were
corrected for this systematic shift.

In the bottom panel of Fig.\ref{fig:ecal} we show the instrument
resolution at 511 keV as a function of the revolution number. Since
the background (internal) 511 keV line is kinematically broadened we
used two bracketing lines (at 438 and 584 keV) to calculate the resolution
at 511 keV. The sawtooth pattern clearly seen in the plot is caused by
the gradual degradation of the SPI resolution due to the detector exposure
to cosmic rays and due to the annealing procedure (around revolution 90)
which restores the resolution. The net result is that the mean resolution near
511 keV is $\sim$ 2.1 keV (FWHM) and over the whole data set the
resolution changes from $\sim$ 2.05 keV to $\sim$ 2.15 keV.

All the data, reduced to the same energy gain, were then stored as 
individual spectra (one per pointing and per detector) with 0.5
keV wide energy bins. These spectra are used for subsequent analysis.

\subsection{Background modeling}
\begin{figure} 
\plotone{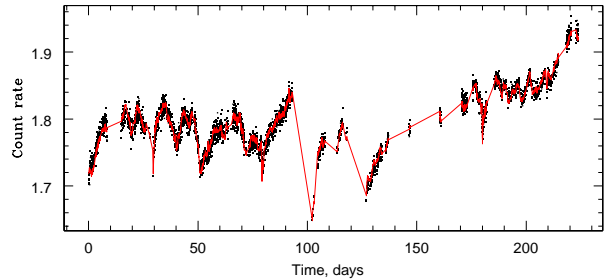}
\caption{Count rate in the 900-1200 keV range (averaged over all 19
SPI detectors) as a function of time. Zero time corresponds to the
beginning of the data set. The curve shows the predicted count rate in
the same band based on a simple background model. 
\label{fig:back}
}
\end{figure}

Once obvious spikes and flares are removed from the SPI data, the
background in the remaining "clean" data is rather stable and it
typically does not vary by more than $\sim$10\%. However the 511 keV
line observed from the GC region produces an excess signal at the level of
1-2 per cent of the background line and therefore variations of the
background have to be taken into account. Ideally one would prefer to
have a background model which is based on some accurately measurable
quantities (like charged particles count rate) so that the background
subtraction does not introduce extra noise to the data. Since no
such model has been provided so far, we use the same data set to build
a provisional background model. When doing so one has to bear in mind
that the statistical significance of the accumulated data is limited (when
narrow energy bins are considered) and the model has to be kept as
simple as possible to provide a robust result. The simplest background
model, which we found acceptable at the present stage of SPI analysis,
assumes that the background is linearly proportional to the Ge detectors
saturated event rate and time. An example of observed and predicted
background for the 900-1200 keV range is shown in
Fig.\ref{fig:back}. Such a broad band was selected to show variations
of the background more clearly. The points in Fig.\ref{fig:back} show
actual measurements (averaged over all 19 SPI detectors), while the line
connects predicted background values. One can see that most of
the prominent background variations are well reproduced by the adopted
model. However, for some revolutions further improvement of the model
is possible once more data, especially blank-field observations,
become publicly available.

For the purpose of the GC data reduction we then regenerated
a background model using all data excluding the central 30\deg (radius)
region centered at GC. The total (dead time corrected) exposure of the
background fields used is 3.7 Msec.

\begin{figure*} 
\plottwo{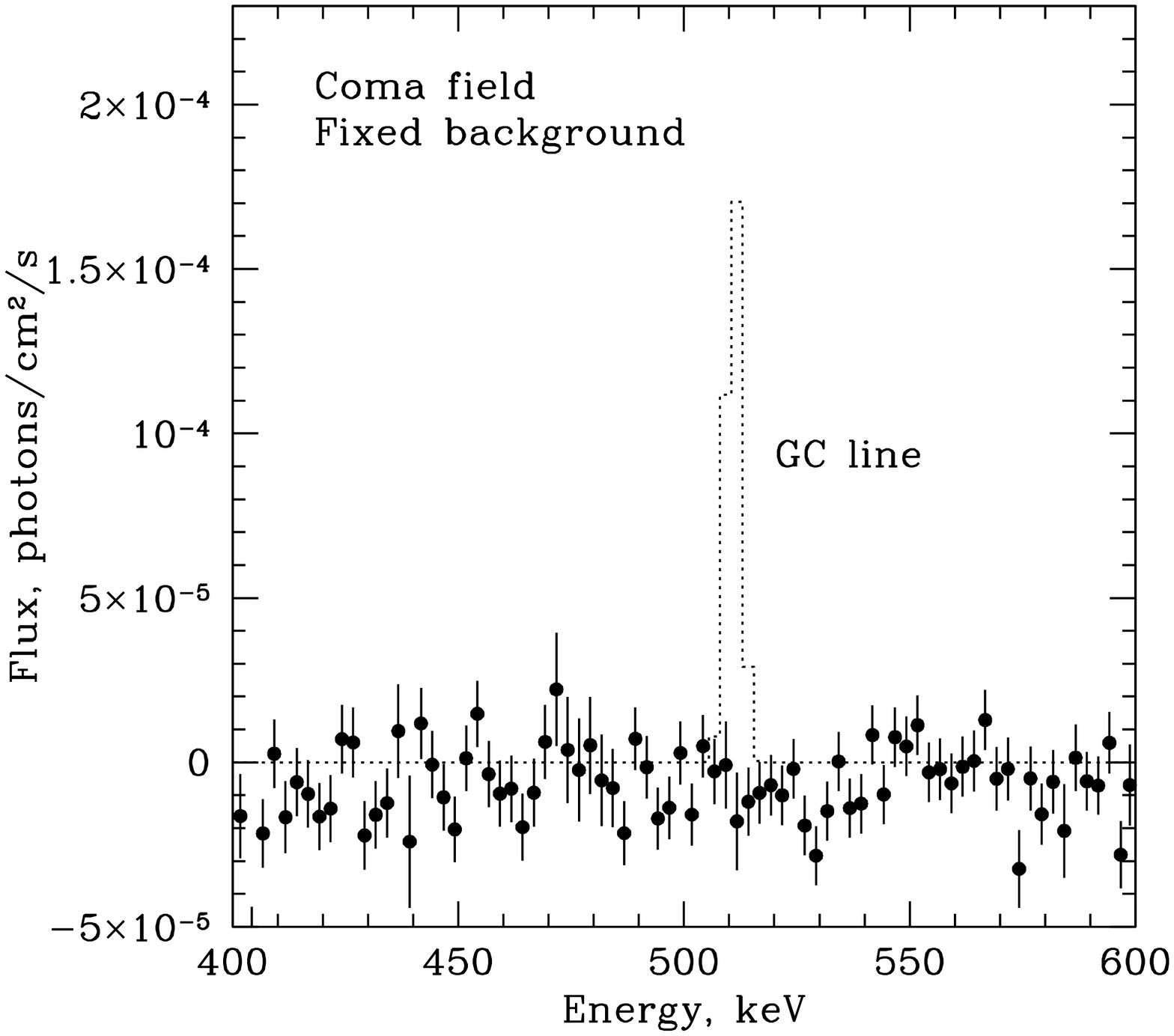}{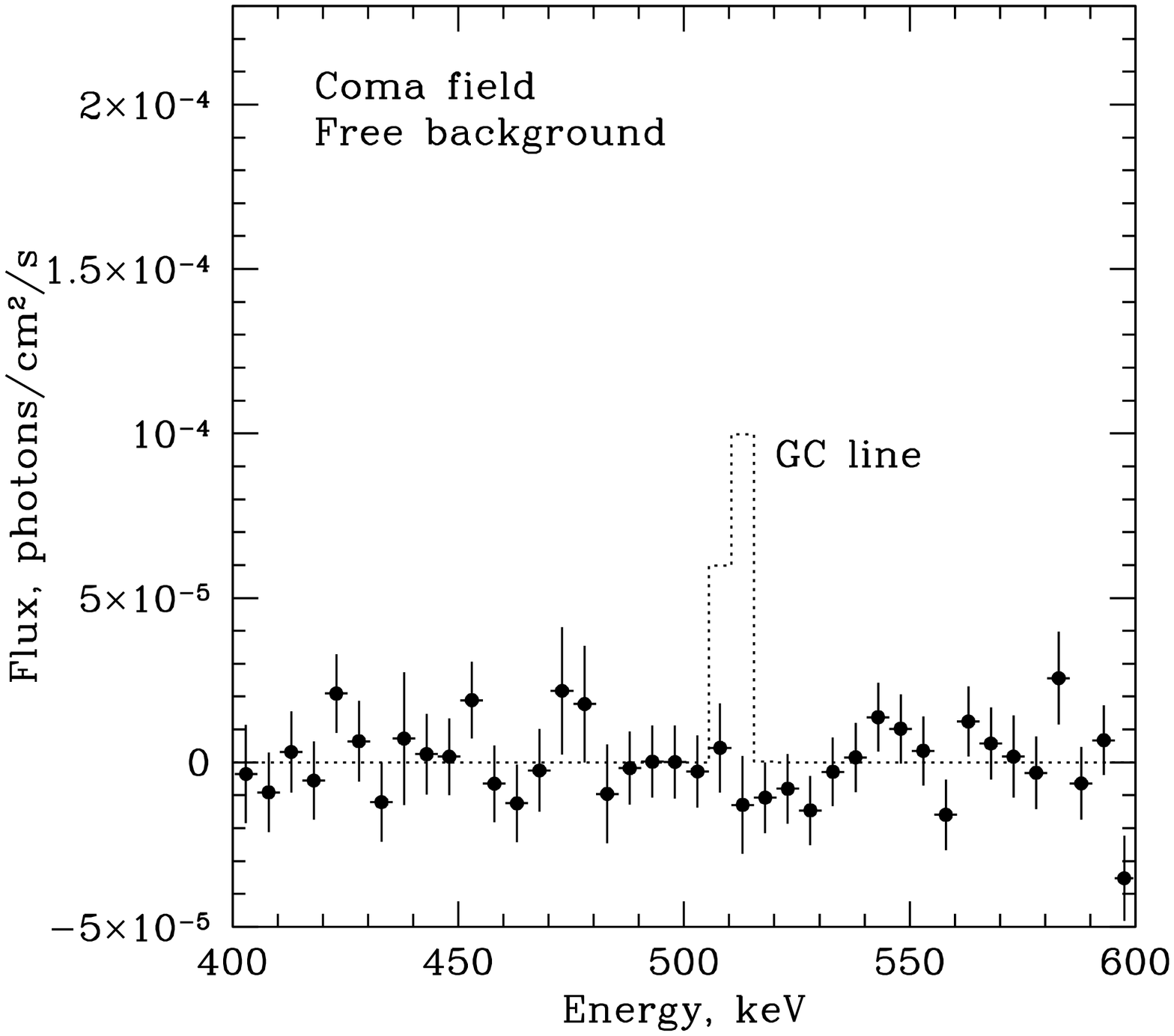}
\caption{Data points show the background subtracted spectra obtained
from the 0.5 Msec SPI observations of the Coma region using the same
procedure as is used for the Galactic Center region. For the left
panel the spatial model is 6\deg Gaussian, centered at Coma cluster,
while the background is fixed. For the left panel the model consists
of a Gaussian plus a constant with the free normalization. For
comparison superimposed on each plot is a Gaussian line at 511 keV
(dotted lines; rebinned to the data resolution) with the total flux as
seen from the GC region. An even $\sim$5 times weaker line would be seen
in this 0.5 Msec observation.
\label{fig:blank}
}
\end{figure*}

To verify the quality of the background model in the energy range of
interest (i.e. around 511 keV) we used 0.5 Msec SPI observations of the
Coma region. While these particular observations were also part of the
data set used for the generation of the background model, they 
contribute only $\sim$10\% to the total background exposure.
Applying exactly the same procedure (described below) used for
the Galactic Center observations we extracted the spectrum assuming
that the spatial distribution of the line flux has a shape of a
Gaussian with FWHM=6\deg centered at the Coma cluster (Fig.\ref{fig:blank}
left panel). As for the Galactic Center observations we used an
additional model consisting of a Gaussian plus constant
(Fig.\ref{fig:blank} right panel). By construction the left plot is
basically the difference between the spectrum coming from the Coma
region and the mean spectrum over the entire background data set which
contains many Galactic Plane pointings with bright sources. Therefore,
a small negative bias, seen in the left panel of Fig.\ref{fig:blank}, is a
natural result. For the right panel (allowing for an additional free
background component constant over all detectors) this negative bias
is not present. For both panels no significant spectral features are
present near the 511 keV line. For comparison we show in
Fig.\ref{fig:blank} a line at 511 keV with parameters similar to those
observed in the GC region. No evidence for spectral features near
511 keV is seen in the Coma field, suggesting that the background is
removed with a sufficient accuracy.

\subsection{Spectra extraction}
For the GC spectra we use the data obtained when the main axis of the
instrument was within 30\deg of the GC direction with an overall
exposure time of $\sim$3.9 Msec (dead time corrected). The spatial model
used is a simple Gaussian with FWHM ranging from 2 to 26\deg. Using
available response files (results of Monte-Carlo modeling, described by
Sturner et al., 2003) the count rate was predicted for every pointing
and every SPI detector. Two models are used to extract the spectra. In
the first model the normalization of the model (in a given energy
band) was then obtained from a simple linear $\chi^2$ fit to the data:
\begin{eqnarray}
\sum_i \left ( \frac{A\times P_i - (D_i-B_i)}{\sigma_i} \right)^2={\rm
min},
\end{eqnarray}
where the summation is over the data set, $A$ is the normalization of the
model (free parameter), $P_i$ is the model predicted rate, $D_i$ is
the observed rate in a single detector during a given pointing, $B_i$
predicted background rate, and $\sigma_i$ is the standard deviation for
the observed rate. In order to avoid bias due to the correlation of the
observed rates $D_i$ and the errors $\sigma_i$ (see Churazov et al.,
1996) we evaluate $\sigma_i$ using the exposure time of the $i-$th
observation and the mean count rate (in a given detector and given
energy band) averaged over a large number of observations. Since the absolute
values of the variations of the count rate are small this procedure
provides an unbiased estimate of $A$ with statistical uncertainty very
close to the theoretical limit. The first model is essentially the
difference between the flux measured from the GC region (weighted with
a spatial Gaussian of a given width) and the mean spectrum over the
data set used for the generation of the background model. For the second
model we allow an additional background component (constant in time and
space, but variable in energy) with a free normalization. I.e. in the
second model the predicted count rate in the $i-$th observation is
$A\times P_i+C$, where both $A$ and $C$ are free parameters. While the
second model has an obvious advantage compared to the first model (see
e.g. Fig.\ref{fig:blank}), especially when trying to detect weak
continuum radiation on top of the strong background line, the addition
of the second free parameter increases (for the present data set) the
statistical error in $A$ by a factor of 1.6 and 2.6 for the 2\deg and
26\deg Gaussian, respectively. We therefore extracted spectra using
both models (hereafter model I and II) and checked the results for
consistency.

The choice of these simple models (with the minimal number of free
parameters) is primarily driven by a desire to get maximum
significance in the resulting spectra. Observations during balloon
flights (e.g. Harris et al., 1998) and with OSSE/CGRO (Kinzer et al.,
2001) and earlier SPI imaging analysis (Kn\"odlseder et al., 2003)
indicate that a Gaussian is a reasonable first approximation for a 511
keV flux excess near GC even though the distribution might be more
complicated than a simple Gaussian. In principle the assumed spatial
distribution mainly affects the absolute normalization of the flux and
to a lesser degree the spectral shape (unless the spectrum shape varies
strongly across the studied area of the sky).

\begin{figure} 
\plotone{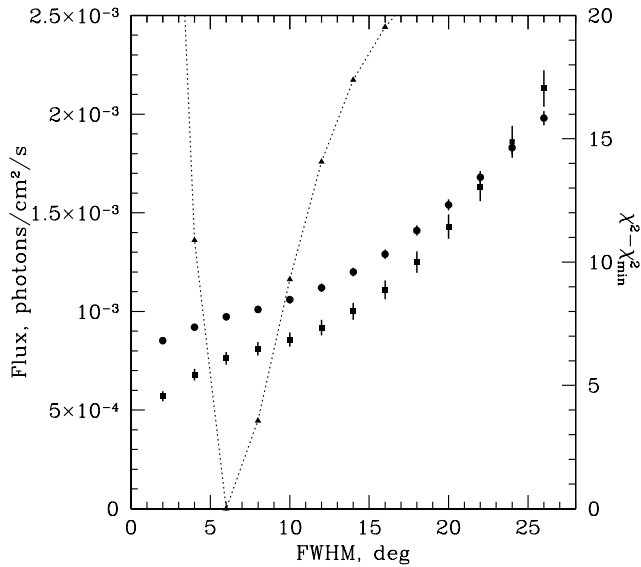}
\caption{Dependence of the flux in the 508-514 keV band on the width
of the spatial Gaussian for model I with a fixed background (solid
circles) and model II with a free constant background (solid
squares). The dotted connecting solid triangles shows the behavior of
the $\chi^2$ in model II.
\label{fig:flux}
}
\end{figure}

The total flux in the 508-514 keV band obtained using both models is
shown in Fig.\ref{fig:flux} as a function of the Gaussian
width. One can see that the flux changes from $\sim0.5~10^{-3} {\rm
ph~cm^{-2}~s^{-1}}$ to $\sim2~10^{-3} {\rm ph~cm^{-2}~s^{-1}}$ as FWHM
changes from 2 to 26\deg. The dotted line shows the behavior of
$\chi^2$ in the model with a free constant (model II) as a function of
the Gaussian width. Within this model the minimum $\chi^2$ is reached
for the FWHM$\sim6$\deg. The absolute value of
the $\chi^2$ for 6\deg Gaussian is 0.9992 per degree of freedom (for
38969 d.o.f.). Given that the S/N for individual observation is very
small, the absolute value of $\chi^2$ is not a very useful indicator
of the acceptability of the model. However its closeness to unity
shows that observed variations of count rates are very close to those
expected from a pure statistical noise. 

The width of the distribution suggested by the above analysis
($\sim$6\deg) is rather close to the value derived for the central
bulge from OSSE observations (Kinzer et al., 2001), while the earlier
analysis of a shorter SPI data set suggested a somewhat broader
distribution $\sim$9\deg (Kn\"odlseder et al., 2003) although
consistent with 6\deg within the quoted uncertainties. The behavior of
the curves in Fig.5 indicates that the 6\deg Gaussian does not account
for the total flux of 511 keV photons coming from the GC region and
an extra component (broader than 6\deg Gaussian) is needed.  This
result is robust against various assumptions on the SPI 
internal background. Since the topic of this paper is the shape of
the annihilation spectrum we will not elaborate on particular
spatial models. Subsequently we use only the spectra
extracted using a 6\deg Gaussian. However, we verified that all the major
spectral parameters (except for the overall normalization) are insensitive
to the width of the spatial model. 


\begin{figure*} 
\plotwide{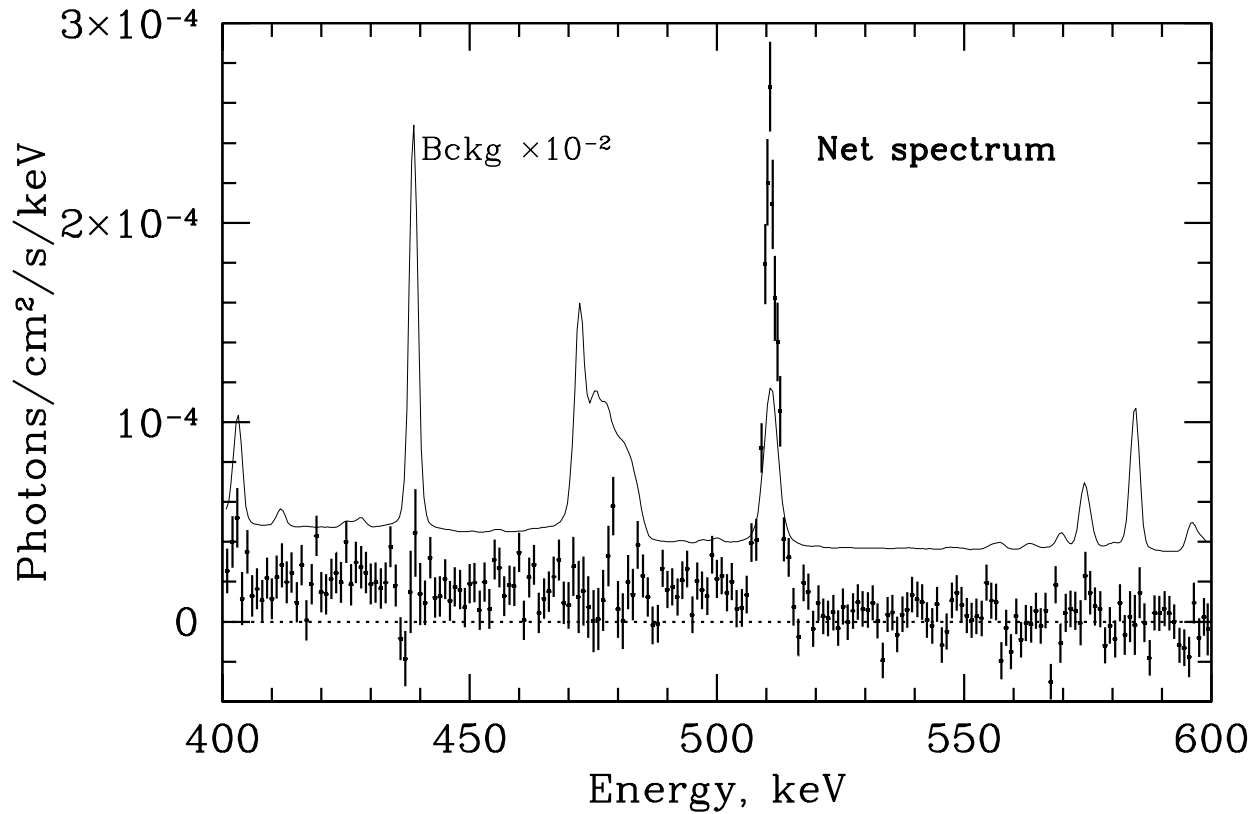}
\caption{Spectrum of the annihilation radiation from the GC region
(6\deg Gaussian) in the 400-600 keV range derived from the model with
the constant background as a free parameter (model II). For comparison
a background spectrum, scaled by a factor of 0.01, is shown with a
solid line. The annihilation line coming from the GC region thus
corresponds to the level of $\sim$ 2-2.5\% of the SPI background 511
keV. The observed continuum radiation below 511 keV is at the level of
0.4\% from the instrument background continuum, except for the
energies close to the positions of the strong background lines where
the useful signal is at the level of 0.1\% of the background.
\label{fig:spback}
}
\end{figure*}

\section{Net spectra and spectral modeling}
\begin{figure} 
\plotone{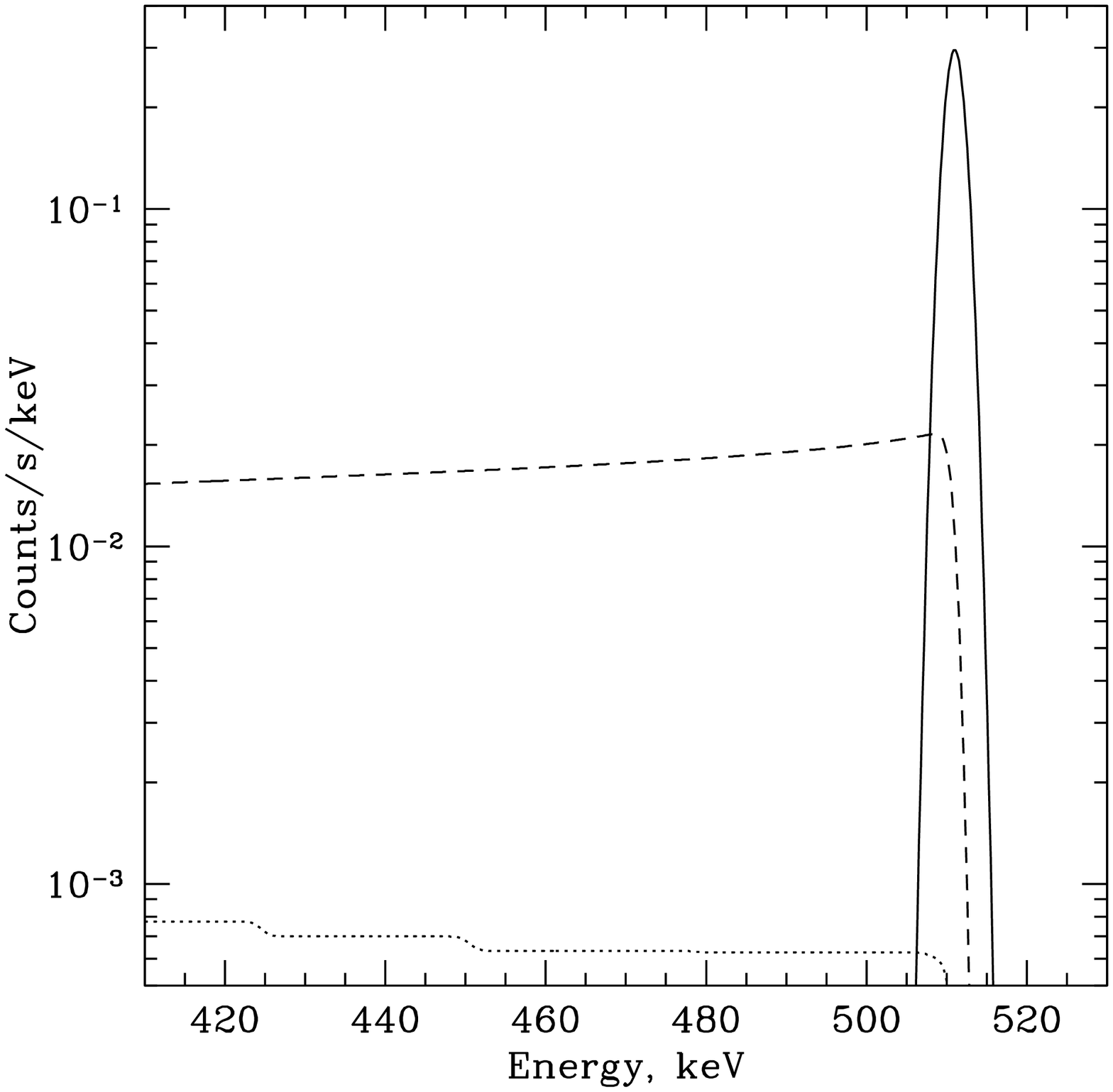}
\caption{Estimated low energy tail (dotted line) of the bright line at
511 keV (solid line). The Monte-Carlo generated response matrix
 (Sturner et al., 2003) was used to estimate the tail. The total
line flux is normalized to unity. For comparison an ortho-positronium
continuum with the total flux 4.5 times larger than the line flux
(i.e. the case of the annihilation through the positronium formation)
is shown with the dashed line.
\label{fig:lowrsp}
}
\end{figure}

The spectra extracted with a 6\deg Gaussian, using two background models
(fixed and free background), are shown in Fig.\ref{fig:spraw} and
Fig.\ref{fig:spback}. The significance of the narrow line detection is
54.1 and 23.8 $\sigma$ respectively (based on the 508-514 energy
band). When fitting the spectra over the energy range near 511 keV it was
assumed that the SPI energy resolution is equivalent to the convolution of
photon spectra with a  Gaussian having FWHM $\sim$ 2.1 keV.  For
fitting spectral lines and the weak continuum on the red side of a line
one has to verify the normalization and shape of the low energy tail
in the SPI spectral response. At present a Monte-Carlo simulated SPI
energy response matrix (\verb+spi_rmf_grp_0003.fits+, Sturner et al.,
2003) is available, which is constructed for broad continuum
channels. According to this matrix the total off-diagonal tail of the
SPI response for $\sim$511 keV photons amounts to $\sim$30\% of the
flux. The tail however is very extended (few hundred keV) and between
450 keV (lower energy boundary used for spectral fitting below) and
511 keV only about 3-4\% of the line flux are due to off-diagonal
response. Shown in Fig.\ref{fig:lowrsp} is a typical situation one can
expect for the GC spectra. The thin solid line shows a Gaussian line
with a total intensity of unity. The dashed line is an ortho-positronium
continuum with the total flux 4.5 times larger than the line flux
(i.e. the case of annihilation through positronium
formation). For comparison the dotted line shows the low energy tail
of the narrow line estimated from the available response matrices. One
can see that indeed the contribution of the tail is very minor (few
per cent relative to the positronium continuum above 400 keV). We however
routinely included this component in the subsequent spectral fitting,
linking its flux to the normalization of the narrow 511 keV line. The
impact of the off-diagonal tail on the continuum is even smaller and we
neglected the tail in the continuum modeling. E.g. for the
positroniuum continuum (the hardest continuum component used in the
model) an account for the tail contribution above 450 keV can change
the normalization by less than 2\%. The flux ratio of the positronium
continuum and the narrow 511 keV line determines the positronium
fraction $F_{PS}=2/(1.5+2.25*F_{2\gamma}/F_{3\gamma})$, where
$F_{2\gamma}$ and $F_{3\gamma}$ are the line and continuum flux,
respectively. For large flux ratios (see Table 1) the positronium
fraction is a weak function of the ratio and unless the Monte-Carlo
simulated off-diagonal response is underestimated by a large factor
(larger than $\sim$5) we do not expect any drastic changes in spectral
parameters.


For the ortho-positronium continuum we use the spectrum of Ore \& Powell
(1949). To allow for broadening of the ortho-positronium
continuum edge at 511 keV (SPI energy resolution and intrinsic
broadening of the edge) we convolved the continuum
with a Gaussian having a width from 2.1 to 7. keV. The fits were found
to be fairly insensitive to the exact value of the edge
broadening (unless it is very large) and instead of introducing an
extra free parameter we fixed the width of the ortho-positronium edge at
4 keV. 

The best fit parameters to the spectra are shown in
Table \ref{tab:fit}. For comparison we show in 
Table \ref{tab:res} the spectral parameters derived by various
missions in the past. It is clear that the results are in broad
agreement. Deep INTEGRAL observations provide the most stringent
constraints on the line centroid and on the width of the 511 keV
line.

\begin{table*} 
\caption{Best fit parameters for the GC spectra calculated for a 6\deg
(FWHM) Gaussian spatial model. Quoted errors are 1$\sigma$ for a single
parameter of interest.}
\begin{tabular}{l l l}
\hline
& Model I & Model II \\
& Fixed Background & Free Background \\
& 485-530 keV & 450-550 keV \\
\hline
&\multicolumn{2}{c}{Single Gaussian + Ortho-positronium + power law} \\
\hline
$E_{1}$, keV & 510.988 [510.95-511.02]     & 510.954 [510.88-511.03] \\
$FWHM_{1}$, keV & 2.47 [2.36-2.58]         & 2.37 [2.12-2.62] \\
$F_{2\gamma}~10^{-4} phot~s^{-1}~cm^{-2}$  & $8.7\pm 0.20$ & $7.16\pm 0.35$\\
$F_{3\gamma}~10^{-4} phot~s^{-1}~cm^{-2}$  & $45.6\pm 4.3$ & $26.1\pm 5.7$\\
$F_{3\gamma}/F_{2\gamma}$ & 5.2$\pm$0.51   & 3.65$\pm$0.82\\
$F_{PS}$                  & 1.035$\pm$0.023 & 0.94$\pm$0.06\\
Power law photon index $\alpha$ & 2.0 (fixed) & 2.0 (fixed)\\
$\chi^2$ (d.of.) & 151.2 (83)              & 192.7 (193)\\
\hline
\hline
\end{tabular}
\label{tab:fit}
\end{table*}

\begin{table*} 
\caption{Parameters of GC $e^+e^-$ annihilation spectrum obtained by
various missions.}
\begin{tabular}{l l l l l}
\hline
Instrument & Line energy (keV) & FWHM$^a$ (keV) & Positronium fraction & References\\
\hline
\hline
GRIS & -- &2.5$\pm$0.4 &  -- & Leventhal et al. 1993 \\
HEAO-3 & 510.92 $\pm$ 0.23 &1.6$\pm$1.35 &  -- & Mahoney et al. 1994 \\
HEXAGONE & 511.53 $\pm$ 0.34 &2.73$\pm$0.75 &  -- & Durouchoux et al. 1993 \\
OSSE & --  & -- &  0.93$\pm$0.04 & Kinzer et al. (2001) \\
TGRS & 510.98 $\pm$ 0.1 &1.81$\pm$0.54 &  0.94$\pm$0.04 & Harris et al. (1998) \\
SPI/INTEGRAL & 511.06 $\pm$ 0.18 &2.95$\pm$0.5 &  -- & Jean et al. (2003) \\
SPI/INTEGRAL & 510.954 $\pm$ 0.075 &2.37$\pm$0.25 &  0.94$\pm$0.06 & this work \\
\hline
\hline
\end{tabular}
\begin{quote}
$^a$ - the errors in FWHM were recalculated from the lower and
upper limits on FWHM quoted in the original papers. 
\end{quote}
\label{tab:res}
\end{table*}

\section{Constraints on the parameters of the annihilation medium}
\begin{figure} 
\plotone{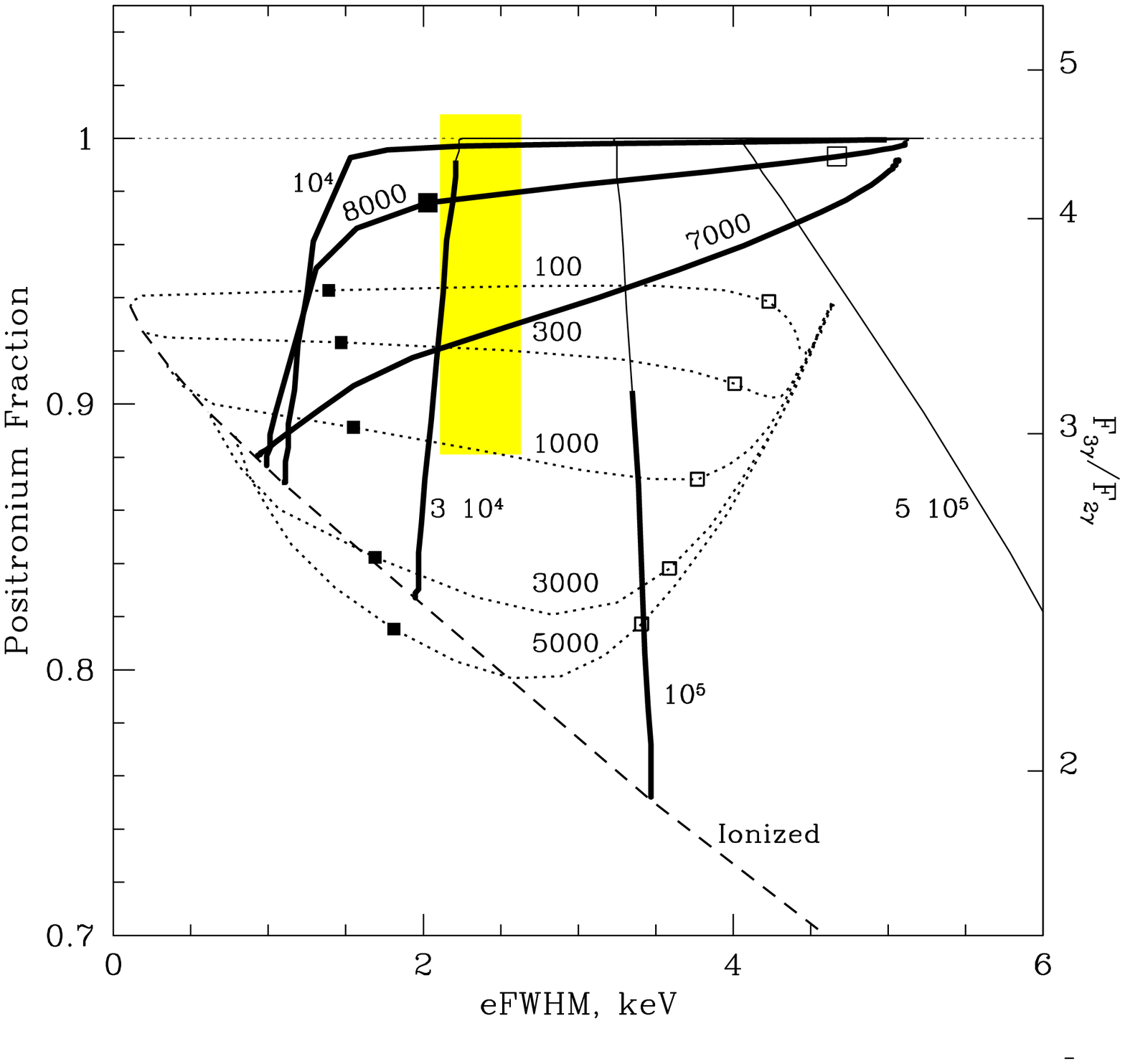}
\caption{The effective FWHM of the 511 keV line versus the fraction of
annihilation through the positronium formation. The gray area is the width
and the positronium fraction observed by SPI. There are two groups of
theoretical curves: cold - $T_e \le 5000$ K (dotted lines) and
warm/hot - $T_e \ge 7000$ K (solid lines). The temperature is fixed
for each curve (the labels next to the curves), but the ionization
fraction varies so that plasma changes from neutral to completely
ionized along the curve. For cool temperature curves and for the 8000
K curve the points corresponding to the ionization degree of 0.01 and
0.1 are marked with the open and solid squares respectively. Each high
temperature curve has two regimes, shown by thin and thick solid
lines respectively. Thin (thick) lines correspond to the ionization
fractions smaller (larger) than expected for collision dominated
plasma at this temperature. For ISM the overionized state (regime
shown by thick lines) is more natural than the underionized (thin
lines). Finally the dashed line shows the relation between the line
width and positronium fraction for a completely ionized plasma as a
function of temperature. 
\label{fig:ofwhm}
}
\end{figure}
Two observed quantities (the width of the line and the strength of the
ortho-positronium continuum) impose some constrains on the temperature
and ionization state of the annihilation medium. The positrons are
assumed to be born hot, with the energy of order of  few hundred keV
and slow down by Coulomb losses (in ionized plasma) and ionization and
excitation (in neutral gas). Once the energy of the positron drops
below few hundred eV, charge exchange reactions with neutrals
and/or radiative recombination or direct annihilation with free or
bound electrons produce the annihilation line and the 3-photon continuum.
The comprehensive description of the slowing down and annihilation of
positrons in ISM is given by Bussard, Ramaty and Drachman
(1979). Subsequent publications (e.g. Guessoum, Ramaty \&
Lingenfelter, 1991, Wallyn et al., 1994, Guessoum, Skibo \& Ramaty
1997, Dermer \& Murphy 2001) used updated values of the cross
sections, but all principle results were in line with the conclusions
of Bussard et al., 1979. For our calculations we considered pure
hydrogen, dust free gas. For ionization, excitation and charge
exchange reactions we used the theoretical cross sections of Kernoghan et
al. (1996), which are in very good agreement with experimental data,
for radiative recombination and direct annihilation with free
electrons the approximations of Gould (1989), and for direct annihilation
with bound electrons the results of Bhatia, Drachman \& Temkin
(1977). For positrons slowing down due to the ionization of hydrogen
atoms, it is necessary to specify the distribution of scattered positrons over
energy. For that purpose we use the distribution over the final
energies predicted by the first Born approximation, while the
normalization (total ionization cross section as a function of the
positron initial energy) was kept at the Kernoghan et al. (1996)
values. One can further correct the distribution over positron energy
losses during ionization taking into account the enhancement of the
cross section when final positron energy is approximately equal to the
energy of the ejected electron (e.g. Mandal, Roy \& Sil, 1986,
Berakdar, 1998). This will be done in subsequent publications. For
positrons slowing down  due to interaction with free electrons we
adopt the analytical approximation of Swartz, Nisbet \& Green
(1971) (see also Emslie 1978) derived for electron energy losses. For
simplicity we used the cold target approximation, i.e. assumed that the
positron energy is much higher than the temperature of the electrons.
In our calculations we assume that charge exchange and radiative
recombination form ortho- and para-positronium according to
their statistical weights, i.e. 3 to 1, and we ignore the tiny contribution of
the direct annihilation to the 3$\gamma$ continuum.

We use the Monte-Carlo approach to trace the history of positrons
deceleration and the formation of the annihilation spectrum taking into
account all the processes mentioned above. The spectrum produced by 
annihilation of thermalized positrons was calculated separately
assuming a Maxwellian distribution of positrons over energy. Our
estimates show that the deviations from the Maxwellian distribution,
resulting from the charge exchange of positrons with neutral atoms,
might affect the results of calculations only for a plasma with a
temperature of $\sim$ 6000 K and a very low ionization fraction (less
than $10^{-3}$), but not for any other combination of temperature and
ionization state.

The predicted annihilation line does not have the shape of a true
Gaussian and often contains a narrow component. The traditional definition
of FWHM would then pick the narrow component and may miss the broader
component. More informative would be the width of an energy interval
which contains a given fraction of the line photons. To simplify the
comparison with the results of previous missions and theoretical
calculations we define the effective full width at half the maximum (eFWHM,
Guessoum et al., 1991) of the line as an energy interval containing
76\% of the line photons. For a true Gaussian eFWHM is equal to FWHM.

The results of the calculations are given in Fig.\ref{fig:ofwhm},
where each curve in Fig.\ref{fig:ofwhm} corresponds to a particular value of
temperature. The ionization fraction varies along each curve so that the
ISM changes from neutral to completely ionized. One can identify
two groups of curves in Fig.\ref{fig:ofwhm} demonstrating
different behavior. 

The first group of curves corresponds to cool gas with a temperature below
$\sim 6000$K. In the cold and neutral ISM about 94\% of the positrons form
positronium in flight, while the rest fall below the positronium formation
threshold of 6.8 eV and eventually annihilate with bound
electrons. The eFWHM of the line produced by positronium formation is 
$\sim$5.3 keV, while the annihilation with bound electrons produces
FWHM of $\sim$1.7 keV (e.g. Iwata, Greaves \& Surko, 1997) due to the
momentum distribution of bound electrons.  The eFWHM in the total
annihilation spectrum is $\sim$4.6 keV. If the ionization fraction in
the cold gas increases above $\sim 10^{-3}$, then Coulomb losses start
to be important and the fraction of positrons forming positronium
in flight decreases. For the positrons falling below 6.8 keV three
processes are important - radiative recombination, annihilation with
free electrons and annihilation with bound electrons.  For ionization
fractions of the order of a few $10^{-2}$ and temperatures $\sim 10^3$ K the
annihilation with bound electrons causes the decrease of the net 
positronium fraction to $\sim$ 90-80\%. If the ionization degree is
more than a few per cent then only radiative recombination and
annihilation with free electrons are important and the positronium
fraction and the line width converge to the values expected for
completely ionized plasma.

The second group of curves corresponds to temperatures higher than
$\sim$7000 K. At these temperatures thermalized positrons can form
positronium via charge exchange with hydrogen atoms and this process
dominates over radiative combination and direct annihilation unless the
plasma is strongly ionized. As a result for  moderate degree of
ionization the positronium fraction is very close to unity. Only when
the plasma is significantly ionized (of the order of 6-10\% for $\sim$8000
K gas and more for higher temperatures) the annihilation with free electrons
starts to be important and the positronium fraction declines with
increasing ionization degree (almost vertical tracks in Fig.\ref{fig:ofwhm}).

The above calculations probe the various combinations of the
temperature and ionization, ignoring the physical possibility of such
combinations. We now compare the results of INTEGRAL observations with
the simulations and discuss plausible conditions.

The observed positronium fraction and the line width are shown in
Fig.\ref{fig:ofwhm} as a shaded rectangle. Comparing the simulated
curves with the results of observations one can conclude that low and
high temperature solutions are possible. The low temperature solution
corresponds to temperatures below 1000 K, while for the high
temperature solution falls into the range from 7000 to $4~10^4$ K. 

\subsection{Positrons annihilating in various ISM phases}
The standard model of ISM in the Milky Way (e.g. McKee \& Ostriker
1977, Heiles 2000, Wolfire et al., 2003) assumes that there are
several distinct phases: Hot ($T_e \ge$ few $10^5$ K), Warm ($T_e \sim
8000$ K) and Cold ($T_e \le 100$ K).

From Fig.\ref{fig:ofwhm} it is clear that hot ($T_e \ge 10^5$ K)
ionized medium does not give a dominant contribution to the observed
annihilation spectrum. Indeed positrons annihilating in such a medium
would produce too small positronium fraction and a much too broad
annihilation line. One can increase the positronium fraction in such
gas by requiring the low ionization fraction (i.e. increase the role
of charge exchange) below the values expected in the collisionally
ionized plasma. Such  situation seems to be very unplausible and the
width of the line remains too large anyway. One can easily limit the
contribution of a very hot ($T_e \ge 10^6$ K), completely ionized
plasma to the observed spectra, by adding a broad line to the
fit. E.g. for $T_e=10^6$K the expected FWHM is $\sim$11 keV (Crannell
et al., 1976) and the 90\% confidence limit on the contribution of
such a line to the observed flux is $\le$ 17\%. Given that at $10^6$K
the direct annihilation and radiative recombination rates are nearly equal
and assuming that the bulk of the remaining line photons are due to
annihilation through the positronium formation, one can conclude that
not more than $\sim$8\% of positrons annihilate in a hot ($>10^6$ K)
medium.

A qualitatively similar conclusion is valid for a cold ($T_e \le$
$10^3$ K) neutral gas. While the positronium fraction is consistent
with the observed one, the width of the line ($\sim$ 4.5 keV) is much
too broad. One can decrease the line width by forcing a large ionization
fraction (more than $10^{-2}$). For cold and dense gas (e.g. in
molecular clouds or cold HI clouds) such an ionization fraction is much
larger than expected.

On the other hand in the warm phase of the ISM ($T_e \sim 8000-10^4$ K)
the ionization fraction varies substantially from less than $\sim$0.1 to
more than 0.8. This phase alone can explain the observed annihilation
line width and the positronium fraction. E.g. for 8000-$10^4$ K plasma
one needs an ionization degree of several per cent to get both
the line width and positronium fraction consistent with
observations. For $2~10^4$ K plasma the ionization fraction has to be
of the order of 0.4. At temperatures higher than $3~10^4$ K even 
plasma in collisional ionization equilibrium is strongly ionized
($>$99\%). In real ISM one can usually expect stronger ionization than
for collisionally ionized plasma. The corresponding parts of the
curves in Fig.\ref{fig:ofwhm} (i.e. ionization fraction larger than
for collisionally ionized plasma) are shown by the thick solid lines.

Given the characteristic shapes of the curves in Fig.\ref{fig:ofwhm} a
combination of annihilation spectra coming from warm ISM and having
various degrees of ionization could produce the annihilation spectrum
consistent with observations.  This conclusion is broadly consistent
with earlier analysis of e.g. Bussard et al., 1979 or Guessoum et
al. 1991.

\subsection{The fraction of positronium formed in flight}
\begin{figure} 
\plotone{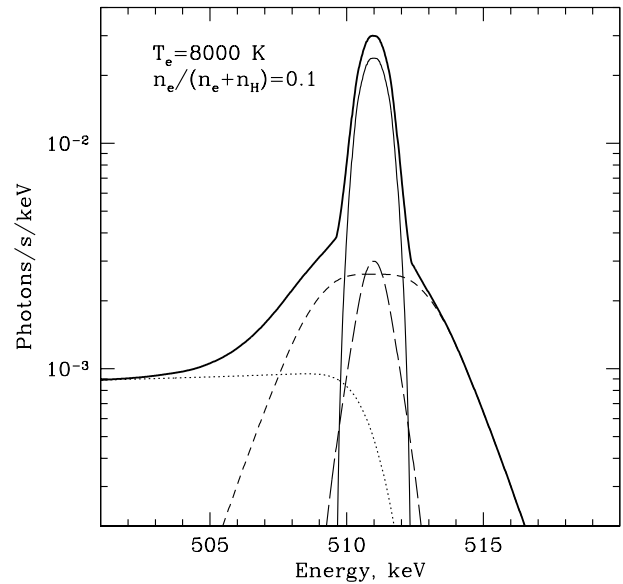}
\caption{The structure of the annihilation spectrum for 8000 K plasma
with the degree of ionization $\sim$ 0.1. Dotted line shows the
ortho-positronium continuum, short-dashed line is the 2-$\gamma$ decay
of positronium formed in flight, thin solid line - 2-$\gamma$ decay
of positronium formed by thermalized positrons, thin long-dashed line
- direct annihilation of thermalized positrons. The thick solid line
shows the total annihilation spectrum.
\label{fig:prof}
}
\end{figure}

\begin{figure} 
\plotone{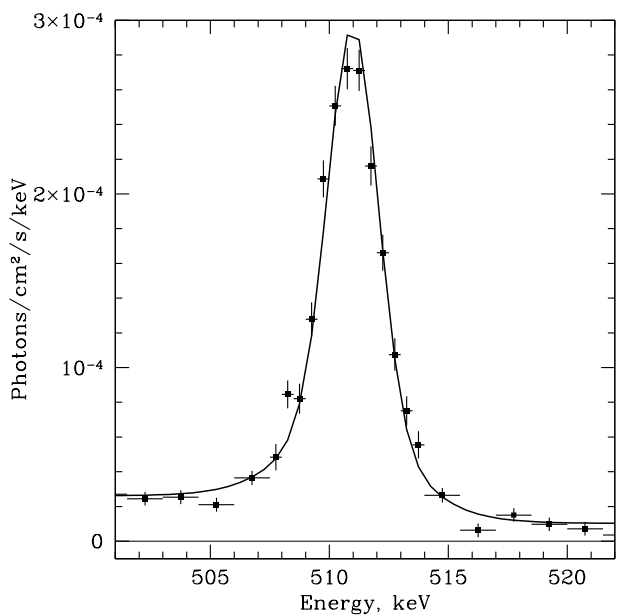}
\caption{The observed spectrum in comparison with a simple model of an
annihilation spectrum of a warm 8000 K medium (ionization fraction
0.1) and a power law. 
\label{fig:wnm}
}
\end{figure}

As was mentioned above, the shape of the annihilation line produced in
the neutral or partly ionized medium can differ from a Gaussian. In
fact for the majority of the theoretical curves in Fig.\ref{fig:ofwhm}
the annihilation line is composed of a rather broad FWHM $\ge$ 5 keV component
(in flight positronium formation) and much narrower FWHM $\sim$
1.0-1.7 keV component (due to radiative recombination with free
electrons, charge exchange of thermalized positrons and annihilation
with bound electrons). An example of the annihilation
spectrum structure is shown in Fig.\ref{fig:prof}. The broad component
(dashed line) is due to in-flight positronium formation, while narrower
components are associated with thermalized positrons. The same model
spectrum (further smoothed with the SPI intrinsic resolution) is
again shown in Fig.\ref{fig:wnm} in comparison with the SPI
spectrum. In terms of the $\chi^2$ such model yields comparable
values, e.g. $\chi^2=192.2$ for 195 d.o.f. for the model with free
background (cf. Table \ref{tab:fit}). 

The change of the line width along the curves is primarily caused by the
variation of the relative weights of these two components. Measurements
of these weights would provide a more powerful test for the ISM
composition than the effective width of the composite line. We
therefore fit the observed spectra with the model consisting of two
Gaussians instead of one as a simplified way to evaluate the
contributions of in-flight and thermalized annihilations. Since for
the in-flight positronium formation 
the width of the annihilation line depends only weakly on other
parameters, we fixed the width of the second Gaussian at 5.5 keV, while
the width of the first Gaussian remained free. The energy of the two
Gaussians were set to be equal. Compared to the single Gaussian model
(see Tables \ref{tab:fit}, \ref{tab:2gau}) this model reduces the
$\chi^2$ by 8.9 and by 3.9 for the ``fixed background'' and ``free
background'' spectra respectively. Given that only one free parameter
(normalization of the second Gaussian) is added to the model, the
F-test suggests that the probability of getting such a reduction in $\chi^2$
is 2.6\% and 5\% for the two spectra, respectively.

\begin{table*} 
\caption{Best fit parameters for the same spectra as in Table 1, when
a second Gaussian with the width fixed at 5.5 keV is added. This second
Gaussian mimics the broad line produced by the in flight positronium
formation. Only essential parameters are quoted.}
\begin{tabular}{l l l}
\hline
& Model I & Model II \\
& Fixed Background & Free Background \\
& 485-530 keV & 450-550 keV \\
\hline
&\multicolumn{2}{c}{Double Gaussian + Ortho-positronium + power law} \\
\hline
$FWHM_{1}$, keV & 1.85 [1.64-2.03]         & 1.50 [1.25-1.89] \\
$F_{2\gamma}~10^{-4} phot~s^{-1}~cm^{-2}$  & $6.63\pm 0.32$ & $5.1\pm0.69$\\

$FWHM_{2}$, keV & 5.5 (fixed)         & 5.5 (fixed) \\
$F_{2\gamma}~10^{-4} phot~s^{-1}~cm^{-2}$  & $2.50\pm 0.67$ & $2.4\pm0.96$\\

$\chi^2$ (d.of.) & 142.35 (82)              & 189.0 (192)\\
\hline
\hline
\end{tabular}
\label{tab:2gau}
\end{table*}

\begin{figure} 
\plotone{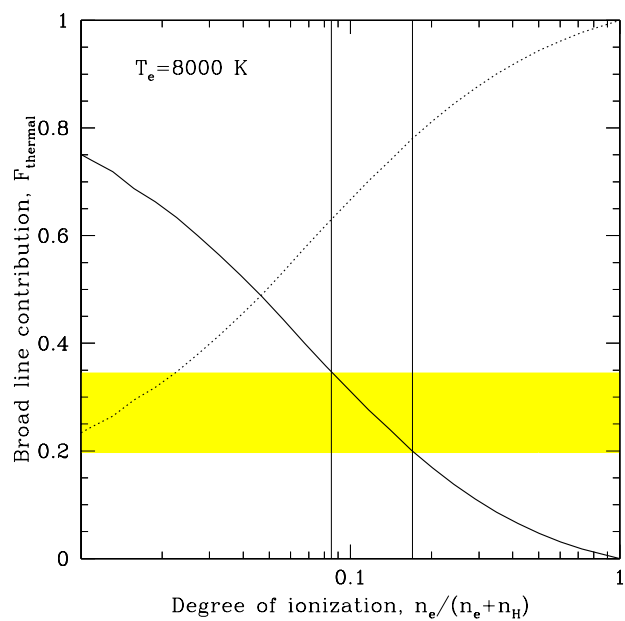}
\caption{Contribution of the positronium, formed in flight, to the
annihilation line (thick solid line) as a function of the ionization
degree for 8000 K plasma. At this temperature thermalized positrons
produce  narrow line, while in flight positronium formation results in
a broad component. INTEGRAL data (2 Gaussians fits) suggest that the
contribution of the broad component is of the order of 30\% (shaded
region). Comparison with theoretical calculations shows that the most
favorable degree of ionization is $\sim$0.1. The dotted line shows the
fraction of thermalized positrons. For 8000 K plasma thermalized
positrons produces a much narrower line than is produced in flight.
\label{fig:ion}
}
\end{figure}

Using 2-Gaussian models one can conclude that the best fit fraction of the
broad Gaussian in the total line flux is $\sim$30\%. One can compare
this value with the expected contributions of the broad (in flight) and
narrow (thermalized) components in the warm medium as a function of
the ionization degree as shown in Fig.\ref{fig:ion}. One can see that
for 8000 K plasma one needs ionization degrees in the range of
0.07-0.17 to have an appropriate relation between the broad and narrow
components. As is mentioned above the curves shown in
Fig.\ref{fig:ion} essentially reflect the changes of the positrons
thermalized fraction as a function of the ionization state.

The same 2-Gaussian models can be used to get more quantitative
constraints on the contribution of the cold neutral phase to the
annihilation budget. The 90\% confidence limit on the contribution of
the broad (5.5 keV wide) Gaussian is $\sim$ 39\%. Let us assume that
this component is due to in-flight positronium formation in a cold
neutral gas. About 6\% of positrons in the cold gas fall below the
positronium formation threshold and annihilate with bound electrons,
thus contributing to the narrower component of the line. The conservative
assumption that the rest of the positrons are thermalized in a warm
($\sim 10^4$ K) strongly ionized ($\ge$50\%) medium and annihilate
through the positronium formation implies an upper limit on the
fraction of annihilations in the cold phase of 45\%.  Thus an
approximately 1:1 mixture of cold neutral and warm ionized phases
would produce a spectrum resembling the one observed with INTEGRAL. If
instead a mixture of cold neutral and warm weakly ionized phases is
considered, then the limits on the cold component contribution are
stronger. If, for instance, the ionization degree of the warm phase is
less than 10\% then the cold component could not provide more than 10\% of the
annihilations. For  warm phase with an ionization degree of 15\% the
limit on the cold component is $\sim$30\%.  Given the uncertainties in
the determination of the mass fractions of various ISM phases
(e.g. Heiles, 2000) the ``natural'' mixture of annihilations in
proportion to the phase masses could also be consistent with the data.

A more detailed analysis of the INTEGRAL data (with an accurate separation
of phases contributions and stringent constraints on the line
components) will become possible once more data are accumulated and
the knowledge of the instrument background and calibration is
improved. More deep observations of the GC region are already planned
and it should be possible to achieve at least a factor of 2-4 longer
exposure of the GC during the next few years.

\subsection{Width of the annihilation lines due to gas motions}
For completeness we estimate the limits imposed by observations on the
motions of the medium where positrons are annihilating.

First of all - a small statistical error on the line centroid achieved
by SPI ($\sim$0.075 keV) corresponds to the uncertainty in velocity of
$\sim$ 44 km/s. The measured line energy coincides (within the
uncertainties) with the unshifted line and therefore there is no
evidence of bulk motion of the annihilating medium with 
velocities larger than $\sim$40 km/s. Note that further improvement in
the line statistics will require an accurate account for the Earth and
Solar system motion relative to the GC since the magnitude of the
velocities is of the same order.

The limits on the gas differential motions are not as strict. Consider
e.g. the envelope of a star or supernova isotropically 
expanding with the velocity $v$. The intrinsically monochromatic line
will then be observed as a boxy spectral feature with the full width
of $1022\times \frac{v}{c}$ keV. One can then estimate the
effective full width as $eFWHM\sim 2.6 \left ( \frac{v}{10^3~{\rm
km/s}} \right )$ keV. Given that the observed width is $\sim$2.4 keV
and for some ISM phases the intrinsic line can be  relatively narrow
$\sim$1-1.5 keV, velocities larger than $\sim 800 ~{\rm
km~s^{-1}}$ can be excluded.

\section{Conclusions}
Deep observations of the Galactic Center region by SPI/INTEGRAL have yielded
the most precise parameters of the annihilation line to date. The
energy of the line is consistent with the laboratory energy with an
uncertainty of 0.075 keV. The width of the annihilation line is
constrained to $\sim$2.37$\pm$0.25 keV (FWHM) and the positronium
fraction to 94$\pm$6\%. Under a single phase annihilation medium
assumption the most appropriate conditions are: the temperature in the
range 7000-40000 K and the degree of ionization ranging from a few
$10^{-2}$ for low temperatures to almost complete ionization for high
temperatures. E.g. annihilation of positrons in one of the canonical ISM
phases - 8000 K gas with an ionization fraction of $\sim$10\% - would
produce an annihilation spectrum very similar to the one observed by
INTEGRAL. Under the assumption of annihilation in a multi-phase medium,
the contribution of a very hot phase ($T_e \ge 10^6$ K) is constrained
to be less than $\sim$8\%. Neither moderately hot ($T_e \ge 10^5$ K)
ionized medium nor very cold ($T_e \le 10^3$ K) neutral medium can
make a dominant contribution to the observed annihilation spectrum.

Further accumulation of the Galactic Center exposure with INTEGRAL and
improvements in the background knowledge and calibration should make
possible detailed fits of the data with the composite annihilation
spectrum. Accurate separation of various components will place tight
constraints on the width and the relative amplitude of annihilation
features formed in flight and after thermalization. It will therefore
be possible to make an accurate census of the distribution of annihilating
positrons over ISM phases. 

Simultaneously accurate information on the spatial distribution of the
annihilation flux (especially in the latitudal direction) can be
efficiently combined with the data on the spatial distribution of
various ISM phases thus constraining the distribution of positron
sources and the transport of positrons in the Galaxy.

\section{Acknowledgements} 

\sloppypar
We would like to thank SPI PIs V. Schoenfelder, G. Vedrenne, J.-P. Roques
and V.L. Ginzburg, V.V. Zheleznyakov, A.M. Cherepashchuk, S.A. Grebenev
and C. Winkler for their support. We are grateful to J.Berakdar,
A.Lutovinov and L.Vainshtein for useful discussions and the referee,
B.~J. Teegarden, for a very helpful report.

This work is based on observations with INTEGRAL, an ESA project with
instruments and science data center funded by ESA member states
(especially the PI countries: Denmark, France, Germany, Italy,
Switzerland, Spain), Czech Republic and Poland, and with the
participation of Russia and the USA.

The dominant part of the GC exposure used in the paper comes from
deep (2 Ms) GC observations carried out as part of the Russian
Academy of Sciences share of the INTEGRAL data.

\label{lastpage}
\end{document}